\documentclass{aa}  
\usepackage{graphicx,color}
\usepackage{txfonts}
\usepackage{verbatim}
\usepackage{natbib,twoopt}
\usepackage[breaklinks=true]{hyperref} 

\hypersetup{
  colorlinks=true,   
  urlcolor=blue,     
  linkcolor=blue     
}
\bibpunct{(}{)}{;}{a}{}{,} 

\makeatletter
\newcommand{\bibnote}[2]{\global\@namedef{#1note}{#2}}
\newcommand{\biblink}[2]{\global\@namedef{#1link}{#2}}
\makeatother

\makeatletter
  \protected\def\stonyslink{%
     \def\hyper@linkstart##1##2{}\let\hyper@linkend\@empty}
  \newcommandtwoopt{\citeads}[3][][]{%
   \href{http://adsabs.harvard.edu/abs/#3}%
        {\stonyslink \citealp[#1][#2]{#3}}
   \biblink{#3}{\href{http://adsabs.harvard.edu/abs/#3}{ADS}}}
 \newcommandtwoopt{\citepads}[3][][]{%
   \href{http://adsabs.harvard.edu/abs/#3}%
        {\stonyslink \citep[#1][#2]{#3}}
   \biblink{#3}{\href{http://adsabs.harvard.edu/abs/#3}{ADS}}}
 \newcommandtwoopt{\citetads}[3][][]{%
   \href{http://adsabs.harvard.edu/abs/#3}%
        {\stonyslink \citet[#1][#2]{#3}}
  \biblink{#3}{\href{http://adsabs.harvard.edu/abs/#3}{ADS}}}
 \newcommandtwoopt{\citeyearads}[3][][]{%
   \href{http://adsabs.harvard.edu/abs/#3}%
        {\stonyslink \citeyear[#1][#2]{#3}}
   \biblink{#3}{\href{http://adsabs.harvard.edu/abs/#3}{ADS}}}
\makeatother

\newcommand{\dif}{\mathrm{d}}

\begin{document} 

   \title{The diurnal Yarkovsky effect of irregularly shaped asteroids}
\author{Yang-Bo Xu\inst{1,2}
	\and
	Li-Yong Zhou\inst{1}\fnmsep\inst{2}\fnmsep\inst{3}
	\and
	Hejiu Hui \inst{4}
	\and
	Jian-Yang Li \inst{5}
}
\authorrunning{Xu et al.}
\offprints{L.-Y. Zhou, \email zhouly@nju.edu.cn}
\institute{School of Astronomy and Space Science, Nanjing University, 163 Xianlin Avenue, Nanjing 210046, China
	\and
	Key Laboratory of Modern Astronomy and Astrophysics in Ministry of Education, Nanjing University, China
	\and
	Institute of Space Astronomy and Extraterrestrial Exploration (NJU \& CAST), Nanjing University, China
	\and
    State Key Laboratory for Mineral Deposits Research \& Lunar and Planetary Science Institute, School of Earth Sciences and Engineering, Nanjing University
	\and
	Planetary Science Institute, Tucson, AZ, USA}
   \date{}

 \abstract{The Yarkovsky effect plays an important role in the motions of small celestial bodies. Increasingly improving observations bring the need of high-accuracy modelling of the effect. Using a multiphysics software COMSOL, we model the diurnal Yarkovsky effect in three dimensions and compare the results with that derived from the widely adopted theoretical linear model. We find that the linear model presents a high accuracy for spherical asteroids in most cases. The ranges of parameters in which the relative error of the linear model is over 10\% are explored. For biaxial ellipsoidal asteroids (particularly oblate ones), the linear model systematically overestimates the transverse Yarkovsky force by $\sim$10\%. The diurnal effect on triaxial ellipsoids is periodic for which no linear model is available. Our numerical calculations show that the average effects on triaxial ellipsoids are stronger than that on biaxial ellipsoids. We also investigate the diurnal effect on asteroids of real shapes and find it be overestimated by the linear model averagely by 16\%, with a maximum up to 35\%. To estimate the strength of Yarkovsky effect directly from the shape, we introduce a quantity ``effective area'' for asteroids of any shapes, and find a significant linear relationship between the Yarkovsky migration rate and the effective area. This brings great convenience to the estimation in practice.  
 } 
  
   \maketitle
%

\section{Introduction}\label{sec:intro}
The Yarkovsky effect is an important non-gravitational effect for meter-sized to kilometer-sized asteroids. Due to the rotation and revolution around the Sun, an asteroid  has an asymmetric temperature distribution on its surface. Such temperature distribution produces asymmetric thermal re-radiation from the asteroid and thus imposes a net recoil force, causing a slow but continuous alteration of the asteroid's orbit. Caused by rotation and revolution, the effect consists of two components, diurnal and seasonal effects.

This effect was named after Ivan Osipovich Yarkovsky, a Polish engineer in Russia who proposed in 1901 such a thermal effect to compensate for the resistance of ``ether'' (refer to \citetads{2005JBAA..115..207B,2006JHA....37...71B} for more information on Yarkovsky and the discovery of his effect). However, it had been forgotten for nearly half a century until \citetads{1951PRIA...54..165O} noticed it when he was studying the motion of small objects orbiting the Sun. In the meantime, \citetads{1952AZh....29..162R} described the same effect without knowing of Yarkovsky's work. \citetads{1987JGR....92.1287R} observationally discovered the Yarkovsky effect from the unexpected variation of the along-track acceleration of the geodetic satellite LAGEOS, and developed a `LAGEOS-taylored' technique for computing the thermal force on a rapidly rotating body. Thus, sometimes the effect was also called Yarkovsky-Radzievskii effect or Yarkovsky-Rubincam effect. 

It becomes a hot topic because it has many implications in the dynamic evolution of the solar system small bodies and the dynamics of near-Earth asteroid, including hazard mitigation for potentially hazardous asteroids (PHAs). \citetads{1998Icar..132..378F} provided a unified description of the Yarkovsky effect in both diurnal and seasonal variants, and discussed some issues in meteorite science relevant to the effect. They proposed that small asteroid fragments in the main belt have enough mobility to reach the resonances with the help of this effect and then be delivered to the Earth-crossing region, which provides an explanation for the fact that the meteorite cosmic ray exposure ages of near-Earth objects are much longer than their dynamical lifetimes. \citetads{1999Sci...283.1507F} calculated the semimajor axis displacements of asteroids between 1 and 10 kilometers in mean radius over their collisional lifetimes, and proposed that the semimajor axis drift may spread the tight clustering of an asteroid family in an observable way which is known later as the V-shape in the distribution of family members. 

Based on previous models \citepads{1976Icar...29...91P,1995P&SS...43..787A}, \citetads{1998A&A...335.1093V,1999A&A...344..362V} developed a complete linear model for the Yarkovsky force acting on spherical asteroids. A general formulation for both diurnal and seasonal effects was provided for any orientation of the spin axis. The main assumption is that the temperature throughout the body does not vary much. \citetads{1998A&A...338..353V} also developed a similar theory for the diurnal force on spheroidal-shaped bodies. Compared with the corresponding force on a spherical body of the same mass, differences up to a factor of three were found.

With the theoretical modelling of the Yarkovsky effect being almost completed, more attentions were paid to the detection of it and its application to the motion of small objects in the inner Solar system such as main belt asteroids, near-Earth objects and Trojans. \citetads{2003Sci...302.1739C} directly detected the effect for the first time in the motion of a near-Earth asteroid, 6489 Golevka, with radar observations. Then the first detection of the effect for main belt asteroids was made by \citetads{2004Icar..170..324N}.

The Yarkovsky effect could be beneficial to the risk mitigation of hazardous near-Earth asteroids (NEA) colliding with the Earth \citepads{2002Sci...296...77S}. For example, an albedo modification technique was designed as a permanently-acting impact risk mitigation of a hazardous NEA \citepads{2010CosRe..48..430H}.

With a large number of asteroids and asteroid families, the main belt provides a good opportunity to investigate the effect. \citetads{2001Sci...294.1693B} demonstrated that the spreading of Koronis family could be caused by the Yarkovsky effect instead of collisions, which resolved the contradiction that the required ejection velocities derived from collision models are much higher than results from impact experiments and simulations. \citetads{2003Icar..166..131T} studied the 7/3 Kirkwood gap and proposed that the short-lived asteroids in it were replenished by its neighbouring asteroids from the Koronis and Eos families. The Yarkovsky-aided transit of asteroids across the 7/3 gap is further investigated in great details recently by \citetads{2020A&A...637A..19X}. Based on the V shape spreading, \citetads{2015Icar..257..275S} make a family classification and estimate their ages.

The Yarkovsky effect also plays an important role in the long-term dynamics of the Trojans of terrestrial planets. \citetads{2005AJ....130.2912S} concluded that primordial Venus Trojans of tadpole orbits would have disappeared by now due to this effect. As for Mars Trojans, \citetads{2005Icar..175..397S} showed that the effect has little influence on the currently known Mars Trojans but is strong enough to destabilize Trojans smaller than 10\,m within 4.5\,Gyr. \citetads{2019A&A...622A..97Z} present the asymmetric stabilities of the prograde and retrograde rotating Earth Trojans due to the Yarkovsky effect and rule out the existence of undetected primordial Earth Trojans. 

The improving accuracy of observations brings the need of more accurate models. Although the linear model developed by \citetads{1999A&A...344..362V} is convenient to use, its assumptions of small variation of surface temperature and circular orbits may be violated in practice. \citetads{1998AJ....116.2032V} developed a one-dimensional numerical model of the seasonal effect on large asteroids, and they found that the linear approximation led to an overestimation by about 15\%. \citetads{2001Icar..149..222S} simulated the diurnal and seasonal effects on a sphere by directly solving the heat equation in three dimensions with a finite difference method, and they found that a stony asteroid on a high-eccentricity orbit drifted in semimajor axis further than on a circular orbit. A similar numerical evaluation of Yarkovsky effect on the eccentricity was reported by \citetads{2002Icar..156..211S}. \citetads{2003Sci...302.1739C} computed the Yarkovsky effect on the asteroid Golevka using a one-dimensional numerical model incorporating its radar-derived shape.  \citetads{2005dpps.conf..171C} proposed a complex one-dimensional numerical model, and illustrated its power by two examples including a tumbling rotator and a binary system. \citetads{2011MNRAS.415.2042R} developed a new complex model called the Advanced Thermophysical Model to take account of the beaming effect caused by the surface roughness and other effects. 

The benefit of high accuracy of these numerical models is limited by their high complexity and low efficiency in computation. The linear model is still widely adopted for its simplicity. Few discussions on the accuracy of the linear model can be found e.g. in \citetads{2012P&SS...60..304S} for the diurnal effect on a spherical body. Based on the general solution of the temperature distribution analytically obtained from the linear model, \citetads{2012P&SS...60..304S} derive the nonlinear solution with an iterative method, and find that the maximum difference between the linear and nonlinear solutions is 13\% in terms of the rate of semimajor axis changing. Using a multiphysics software COMSOL, \citetads{basart2012modeling} and \citetads{groath2014computational} estimated the Yarkovsky effect on the asteroid 101955 Bennu (provisional designation 1999 RQ36), considered it as a spherical body, and made a simple comparison with the results from \citetads{1999A&A...344..362V} and \citetads{2012P&SS...60..304S}. {}

Considering the great convenience of the linear model of the Yarkovsky effect in application but few discussions on its accuracy, especially for asteroids with irregular shapes, it is worthwhile studying the applicability of the linear model in greater detail. 
Here in this paper we report our investigations on the applicability of the linear model for the diurnal effect by comparing it with the numerical model. Our study concerns two issues. One is the difference between the linear model and numerical simulations for spheroidal asteroids, and the other is the deviation caused by approximating irregular asteroids as spheroids and treating them by the linear model. The rest of this paper is organized as follows. In Sect.~\ref{sect:method}, we review the basic theory of heat transfer problem and set up our numerical model. Then, in Sect.~\ref{sect:compare} we compare the linear model with the numerical simulations, investigating the dependence of validity of the linear model on thermal, dynamical and geometrical parameters of asteroids. Finally, conclusions are summarised in Sect.~\ref{sect:conclusion}.

\section{Method} \label{sect:method}
\subsection{Basic theory} \label{sect:theory}
The Yarkovsky effect is a manifestation of the nonuniform temperature distribution on an asteroid's surface. Therefore, the core issue in modelling the Yarkovsky effect is to derive the temperature distribution in a heat transfer problem. 

The heat transfer happens in two ways in an asteroid, i.e., radiation on the surface and  conduction in the interior. In a rotating body-fixed frame, the conduction can be described by the Fourier equation of temperature $T$:
	\begin{equation} \label{eq:fourier}
		\rho C \frac{\partial T}{\partial t} = K\nabla^2T,
	\end{equation}
where $\nabla^2$ is the Laplace operator, $K, C, \rho$ are the thermal conductivity, specific heat capacity and density of the material, respectively. 
The radiation happens on the asteroid's surface, which leads to a boundary condition reflecting the conservation of energy:
	\begin{equation} \label{eq:boundary}
		\epsilon\sigma T^4 + K \left(\mathbf{n}\cdot \frac{\partial T}{\partial \mathbf{r}}\right) = \alpha \mathcal{E}.
	\end{equation}
The first term on the left-hand side and the right-hand side are radiation terms, accounting for the energy re-radiated by the asteroid to outer space and the energy absorbed from the Sun, respectively. The second term accounts for the energy conducted to the deeper layers of the asteroid. As for the notations, $\epsilon$ denotes the emissivity, $\sigma$ the Stefan-Boltzmann constant, $\mathbf{n}$ the unit vector normal to the surface, $\alpha$ the absorption coefficient and $\mathcal{E}$ the external (solar) radiation flux. These two equations are difficult to solve analytically and some approximations must be made, e.g., the linear approximation proposed by \citetads{1998A&A...335.1093V}. 

As long as the temperature distribution is derived somehow, and assuming a Lambert surface, the net recoil force of the thermal radiation of the asteroid can be obtained:
	\begin{equation} \label{eq:force}
		\mathbf{F} = -\int \frac{2}{3c}\epsilon\sigma T^4 \mathbf{n} \, \dif S,
	\end{equation}
where $c$ is the speed of light, $S$ the unit surface area, and the integral domain is the whole surface.

\subsection{Numerical setup} \label{sect:model}
The temperature distribution in our model is numerically simulated using the commercial software, COMSOL Multiphysics$^\circledR$ \footnote{COMSOL Multiphysics$^\circledR$ v. 5.6. \url{www.comsol.com}. COMSOL AB, Stockholm, Sweden.}, designed for solving complex physical problems involving coupled physical processes in high degree of freedom. Unlike in most 1D numerical models, the interactions among surface elements, including surface-to-surface shading and radiation, are considered in our simulations.

The model is set in three dimension (3D). The heat transfer inside the body by conduction and on the surface by radiation are calculated taking into account the surface-to-surface radiation. A temporal (not stationary) distribution of surface temperature in the rotating body-fixed frame is calculated using COMSOL.  

Assuming the asteroid is a perfect sphere, we create a sphere of radius $R=1\,$m and set its physical parameters. For simplicity, the thermal conductivity $K$ and specific heat capacity $C$ are set as constants, and the emissivity equals the absorption coefficient, $\epsilon=\alpha=0.9$.
Initially the asteroid is assumed to be an isothermal sphere, with the initial temperature determined by the equilibrium between its thermal radiation and absorption from the solar radiation, i.e., $4\pi R^2\cdot\epsilon\sigma T^4=\pi R^2\cdot\alpha \mathcal{E}$. 

The heat flux $\mathcal{E}$ depends on the distance between the asteroid and the Sun. In a rotating body-fixed frame, the Sun orbits the asteroid, acting as an external radiation source. The orbital period of the Sun in fact is the rotation period of the asteroid $P$. 

Observing from an inertial (non-rotating) frame, we are supposed to see finally a static temperature distribution in the asteroid. When the temperature variation at the subsolar point on the surface is less than 1\% in a rotation period, we regard the equilibrium is reached, which generally takes an integration time of $50\,P$ in most cases of our simulations. 

One of the critical steps in numerical scheme of finite element method (as in COMSOL) is to build the mesh. An appropriate mesh must be a compromise between accuracy and efficiency. After plenty of tests for the spherical asteroid model, we generate prisms mesh to fill the spherical shell and stuff the central hollow with tetrahedral mesh. 
The core of sphere barely affects the final temperature on the surface, because the heat flux cannot penetrate deeply in most cases when the object is not too small, thermal inertia not too high, and rotation not too slow. Therefore in practice, a hollow sphere, instead of a solid one, is accurate enough to model the asteroid if a boundary condition of thermal insulation is set on the inner surface of the spherical shell with an appropriate thickness. 

\subsection{Mesh building} \label{sect:mesh}

The heat transfer and temperature variation in an asteroid actually occur mainly on the surface and in the shallow subsurface region, but not deep into the body. We show this by calculating the temperature in the body using a 1\,m-radius asteroid as an example. We parametrize the asteroid like a typical regolith-covered asteroid with a density $\rho=$ 1500\,kg\,m$^{-3}$, a conductivity $K=$ 0.0015\,W\,m$^{-1}$\,K$^{-1}$, and a specific heat capacity $C=$ 680 J\,kg$^{-1}$\,K$^{-1}$. A rotation period $P= 1000$\,s, a spin axis obliquity $\gamma=$ 0$^\circ$, and a heliocentric distance $r=$ 1\,AU are also assigned. These parameters are adopted in the rest of the paper unless noted otherwise. We realize that the period of 1000\,s is really short even for a 1\,m-radius small asteroid, but this value does not affect the following analyses in this paper and in fact the value $P$ can be traded with $\rho$ and $C$, as we will see later. 

\begin{figure}[htbp]
   \centering
   \resizebox{\hsize}{!}{\includegraphics{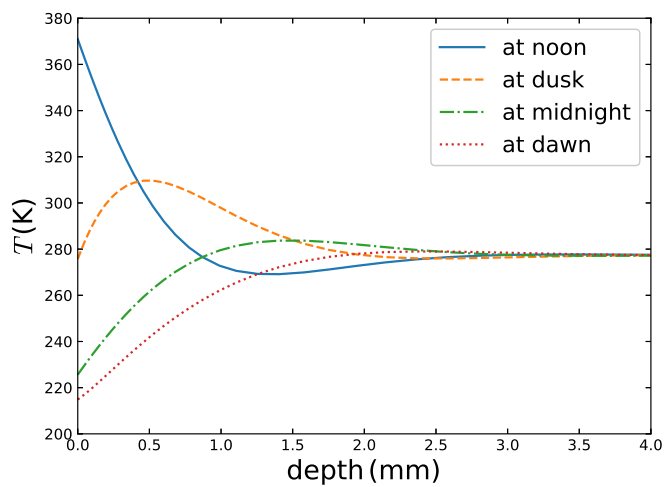}}
      \caption{Temperature along a certain radius on equatorial plane at four different time. The depth is measured downward from the surface.
                       }
         \label{fig:T_depth}
   \end{figure}

As illustrated in Fig.~\ref{fig:T_depth}, along the depth direction, the temperature on each point varies with time as the asteroid rotates and the heat propagates inside. For a point on the asteroid surface (depth $0.0$\,mm) the temperature is high near local noon (when it's the subsolar point), and as the asteroid rotates the temperature drops down through dusk, midnight and dawn. It is worth noting that the hottest moment generally is not at noon but shortly after noon due to the ``thermal inertia'' of the asteroid material. For the same reason, on a subsurface point, e.g. at depth of $0.7$\,mm, the temperature around dusk is higher than at noon. Going deeper inside, the temperature approaches a constant value, i.e. below a certain depth the asteroid core has a static temperature. In practice, for the purpose of studying the thermodynamics, we can simulate the asteroid by a spherical shell of a certain thickness and a solid core of a constant temperature, as we mentioned above in Section~\ref{sect:model}.  

Since the heat transfer and temperature variation happen mainly in the skin layer of the asteroid, the thickness of such layer could be a good reference to our mesh building strategy. In fact, a characteristic quantity called ``penetration depth'' in the literature that can be derived by dimensional analysis is defined as \citepads{1948BAN....10..351W}:
	\begin{equation}
		l_s = \sqrt{\frac{KP}{2\pi\rho C}}.
		\label{eq:pendep}
	\end{equation}
Taking the parameters assigned to the asteroid adopted in this paper, the penetration depth can be calculated $l_s\approx 0.5$\,mm. At this depth, the fluctuation of temperature decreases to $1/e$ of the surface amplitude. Theoretically, at a depth of 9 times $l_s$, the variation reduces to $1/e^9\approx 10^{-4}$. 

\begin{figure}[htbp]
   \centering
   \resizebox{\hsize}{!}{\includegraphics{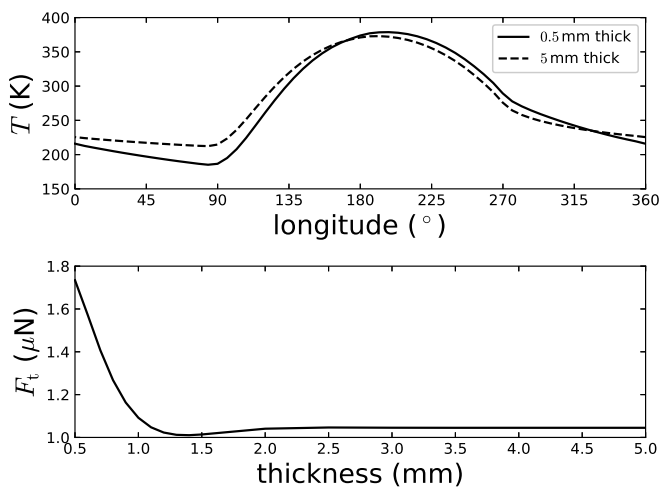}}
      \caption{Influence of adopted shell thickness on the simulated results. Upper panel: Temperature distribution along the equator, for two values of shell thickness. Lower panel: Dependence of the simulated transverse Yarkovsky force on the adopted shell thickness. 
                       }
         \label{fig:F_thickness}
   \end{figure}

To determine a practical thickness of the hollow shell, a couple of tests have been performed. In these tests, we try different thickness of the shell and calculate the temperature on the surface, with which the Yarkovsky force can be obtained through Eq.~\eqref{eq:force}. Since the force component in the transverse direction (perpendicular to the incident solar radiation in the orbital plane) dominates the Yarkovsky effect on semimajor axis drifting, we show the temperature profile along the equator and the transverse Yarkovsky force $F_t$ in our tests in Fig.~\ref{fig:F_thickness}. 

As shown in Fig.~\ref{fig:F_thickness}, the adopted thickness of the presumed shell influences the results significantly. The temperature profile of the case with shell thickness $0.5$\,mm ($=l_s$) is apparently different from the one of thickness $5.0$\,mm (upper panel), which results in a significant difference (about 70\% relatively) in terms of the transverse Yarkovsky force (lower panel). The calculated force varies greatly first as small thickness is adopted, and it gradually converges to a certain value when the thickness increases. Obviously the variation is an artificial effect due to the fact that the adopted shell is too thin to practically embody the thermodynamic cycle of the surface layer within. We find in the lower panel of Fig.~\ref{fig:F_thickness} that the deviation of $F_t$ from the convergent value ($1.04497\,\mu$N) at a thickness of $2.5$\,mm is about $0.1\%$ and it reaches $\sim$10$^{-5}$ at $4.0$\,mm. So in this paper we conservatively set ten times of penetration depth (10$l_s$) as the thickness of the hollow shell, though we will suggest $5l_s$ for general purposes. From now on, the numerical model of an asteroid is assumed to consist of a shell of thickness $10l_s$ and an isothermal core inside whose temperature is the same as on the inner boundary of the shell.

Such mesh strategy leads to tens of thousands of mesh elements under which our desktop workstation computer takes typically one hour to finish the computation for each simulation case till a steady temperature distribution is reached.

\section{Comparisons with linear models} \label{sect:compare}

It is difficult to obtain analytically the temperature distribution on the surface of an asteroid from Eq.~\eqref{eq:fourier} and Eq.~\eqref{eq:boundary}. However, in the linear model, a solution has been found after assuming that the temperature does not vary greatly throughout the asteroid's surface thus the term $T^4$ in Eq.~ \eqref{eq:boundary} could be linearized via Taylor expansion.
But in fact, in the example shown in Fig.~\ref{fig:T_depth}, the temperature varies between a minimum of 215\,K and a maximum of 370\,K, more than 70\% higher than the minimum. In addition, even if the linear model is used, the equations can only be solved analytically for asteroids with certain specific regular shapes. 

Therefore, pure numerical simulations are still necessary in many circumstances. Meanwhile, the linear model provides appropriate estimation of the Yarkovsky effect when the shape of asteroid is unknown. It is worth assessing the deviation of the linear model from the numerical simulations for asteroids with different shapes, which serves as the purpose of this section. We will present our investigations on Yarkovsky effect for spherical, ellipsoidal and irregular asteroids.

\subsection{Spherical asteroid}

Assume a spherical asteroid of radius $R=1.0$\,m on a circular orbit at a distance $r=1.0$\,AU from the Sun, and use the numerical simulation (via COMSOL) and the linear model \citepads{1998A&A...335.1093V}, we calculate the Yarkovsky effect for different values of parameters. By these calculations, we can check the dependence of the Yarkovsky effect on parameters, and verify the validity of the linear model under different conditions. For this purpose, we calculate the Yarkovsky force or the rate of semimajor axis migration.  

As mentioned in Sect.~\ref{sect:mesh}, the heat conduction is limited in a thin layer (we adopt $10l_s$ in this paper). The temperature distribution on an asteroid's surface (thus the accuracy of the linear model) is generally independent of the asteroid's size unless the penetration depth is comparable to the size. Thus we note that the radius of $R=1.0$\,m adopted here is large enough to show the relative accuracies of the linear theory. 

\subsubsection{Dependence on $\rho, C$ and $P$}
\begin{figure}[htbp]
   \centering
   \resizebox{\hsize}{!}{\includegraphics{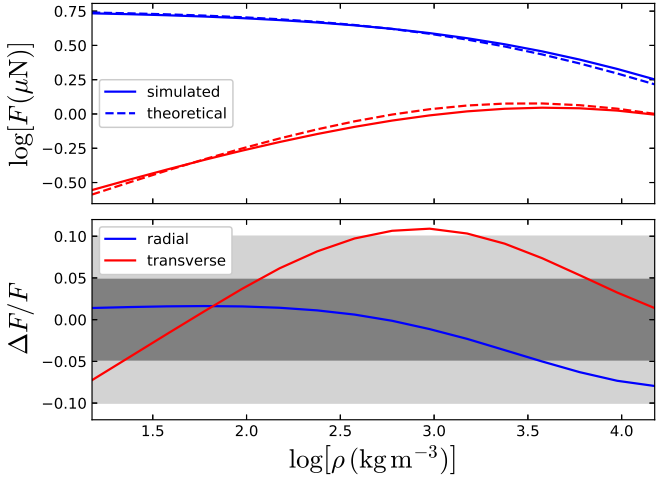}}
      \caption{Comparisons of the Yarkovsky forces derived from the linear model (theoretical) and numerical simulations for varying density $\rho$. Upper panel: The forces (blue for radial and red for transverse components) versus $\rho$. Solid curves refer to simulated results while dashed curves are for the linear model. Lower panel: The fractional differences between results from two methods. Blue and red curves are for the radial and transverse components, while the grey and light grey areas indicate the difference less than 5\% and 10\%, respectively.
                       }
         \label{fig:compare}
   \end{figure}

In Fig.~\ref{fig:compare} we show our calculations with density $\rho$ as the variable. To obtain the results in this figure, all parameters but $\rho$ are fixed ($C= 680\, \text{J kg}^{-1}\text{K}^{-1}$, $K= 0.0015\,\text{W m}^{-1}\text{K}^{-1}$, $P= 1000$\,s as mentioned before). The heat transfers slower when the material density is higher, and the hottest spot on the asteroid surface differs further from the subsolar point. As a result, the transverse component of the Yarkovsky force, which is the net recoil force of the thermal radiation, will increase as $\rho$ increases, while the radial component decreases for the same reason, just as we can see from Fig.~\ref{fig:compare}. In the meantime, the peak temperature decreases with increasing $\rho$, and this is why the transverse force begins to decrease when $\log\rho > 3.5$. 

Similar argument can be made too for the parameter of specific capacity $C$. In fact, if we regard the rotation period $P$ as a parameter and measure time in $P$, i.e. substitute time $t$ by $\tau=t/P$, then Eq.~\eqref{eq:fourier} can be rewritten as
	\begin{equation} \label{eq:fourieral}
\frac{\rho C}{P} \frac{\partial T}{\partial \tau} = K\nabla^2T\,.
	\end{equation}
We note that $\rho, C$ and $P$ appear only here in equations describing the thermal dynamics. Consequently, these three parameters degenerate into a combination $(\rho C/P)$, and thus the Yarkovsky effect depends on $C$ and $1/P$ in a similar way as on $\rho$ (Fig.~\ref{fig:compare}). This has also been confirmed by our numerical simulations.  

The differences between the results from numerical simulations and the linear model can be seen in Fig.~\ref{fig:compare}, particularly in the transverse component of the force that drives the asteroid to drift in semimajor axis. It must be noted that the differences are acceptable. The relative difference shown in the lower panel of Fig.~\ref{fig:compare} is nearly always within 10\%, implying that the linear model works very well in most cases for spherical asteroids. 

We note that the relative difference for transverse force is a little above 10\% for the density around $\sim$10$^3$\,kg\,m$^{-3}$. Taking the fixed parameters $C$ and $P$ adopted in calculations, this corresponds to $\rho C/P\sim 680\,\text{J K}^{-1} \text{m}^{-3} \text{s}^{-1}$. In addition, at around $\log\rho=1.7$ corresponding to $\rho C/P\sim 35\,\text{J K}^{-1} \text{m}^{-3} \text{s}^{-1}$, the relative deviations of both radial and transverse forces are very close to zero simultaneously. If we fix $\rho=1500$\,kg\,m$^{-3}$ and $C= 680\, \text{J kg}^{-1}\text{K}^{-1}$, these two $\rho C/P$ values can be translated to rotation period of 1\,500\,s and 30\,000\,s, respectively. 

\subsubsection{Dependence on $K$}

\begin{figure}[htbp]
	\centering
	\resizebox{\hsize}{!}{\includegraphics{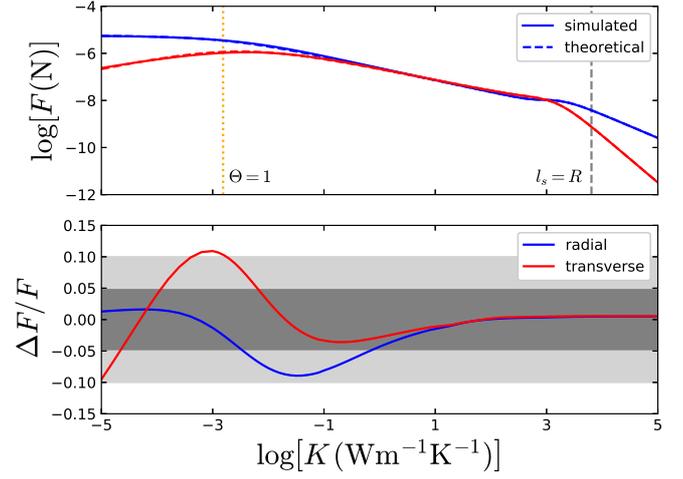}}
	\caption{Same as Fig.~\ref{fig:compare} but for varying thermal conductivity $K$. In the upper panel, the vertical orange dotted line refer to the location of $\Theta=1$ while the vertical grey dashed line indicates where the penetration depth equals the asteroid's radius. 
	}
	\label{fig:K}
\end{figure}

Different from $\rho, C$ and $P$, the thermal conductivity $K$ is involved in both Eq.~\eqref{eq:fourier} and Eq.~\eqref{eq:boundary}, which might make the dependence on it more complicated. Meanwhile, being difficult to measure directly in astronomical observation, $K$ of an asteroid is generally of high uncertainty. We make the similar calculations for varying $K$ as we have done for $\rho$, and summarize the results in  Fig.~\ref{fig:K}. 

In general, a large value of conductivity $K$ is conducive to the heat transfer inside an asteroid, so that the temperature on its surface tends to reach a uniform distribution quickly, leading to a small Yarkovsky effect, which is reflected by the overall descending profile of curves in the upper panel of Fig.~\ref{fig:K}. 

To describe the relaxation of surface temperature with respect to the rotation, \citetads{1998Icar..132..378F} introduced a parameter $\Theta$, which is defined as the ratio between the infrared energy radiated away from the surface in a rotational period ($\approx 4\pi R^2\epsilon\sigma \bar{T}^4 P$) and the thermal energy stored in the ``thermally-affected surface shell'' ($\approx 4\pi R^2\delta\rho C \bar{T}$):
	\begin{equation} \label{eq:theta}
\Theta=\frac{4\pi R^2\delta\rho C \bar{T}}{4\pi R^2\epsilon\sigma \bar{T}^4 P} = \frac{\rho C \delta} {\epsilon\sigma P \bar{T}^3}\,,
	\end{equation}  
where $\bar{T}$ denotes the average temperature in the ``thermally-affected surface shell'' with a thickness of $\delta$, which can be roughly estimated as the penetration depth $\delta = l_s$. According to the definition, $\Theta>1$ indicates that energy would be stored in the surface layer efficiently while the temperature gradients on the surface would be built up inefficiently, resulting in a small Yarkovsky effect, and {\it vice versa}. 

Three regimes can be seen in the upper panel of Fig.~\ref{fig:K}. From left to right, in the first regime of small $K$, the radial component of the Yarkovsky force is higher than the transverse component, implying that the radiation delay angle (the angle between the hottest spot and the subsolar point) increases with $K$ but is always beneath $45^\circ$.  This regime is expected to be separated by $\Theta=1$ from the second regime where the radial Yarkovsky force is almost the same as the transverse force.  

In the second regime, the transverse Yarkovsky force being approximately equal to the radial force implies a radiation delay angle of almost $45^\circ$. This regime ends when the thickness ($\delta$) of the thermally affected layer equals the asteroid radius. Obviously, the thickness $\delta$ increases with $K$, thus for a given radius $R$ the condition $\delta = R$ is met at certain value of $K$. We plot the location of $\delta=l_s =R$ in Fig.~\ref{fig:K}. It is worth to note that two lines of $\Theta=1$ and $l_s = R$ serve as the boundaries of three regimes fairly well. 

In the third regime where $l_s>R$, the radial force deviates from transverse force, i.e., the delay angle begins to decrease from $45^\circ$. When an asteroid is so small (or equivalently, the conductivity is so large) that the heat may transfer easily across the body from one side to the opposite side, the temperature gradients (even between the day and night sides) will be hardly built and the Yarkovsky effect begins to vanish quickly. 

As shown in the lower panel of Fig.~\ref{fig:K}, the relative difference between the numerical simulation and linear model is always within 10\%, and it converges to zero with increasing $K$. For large thermal conductivity, the energy from solar radiation can be quickly conducted to other regions, and the variation of temperature across the asteroid will be small so that the condition for linear model is very well satisfied. But the difference does not vanish monotonically with increasing $K$ because the Yarkovsky force depends not only on the temperature asymmetry on the surface but also on the average temperature. 

Substitution of the thickness $\delta$ in Eq.~\eqref{eq:theta} by the penetration depth in Eq.~\eqref{eq:pendep} leads to $\Theta = \sqrt{\rho C K}/(\epsilon\sigma T^3 \sqrt{2\pi P})$. We note that $\rho, C, K$ in the numerator are all parameters about the material. In fact, the numerator is just the ``thermal inertia''
\begin{equation} \label{eq:inertia}
\Gamma = \sqrt{\rho C K}\,,
\end{equation}
which represents the capacity of a material to store heat and to delay its transmission \citepads{ferrari2007building}. $\Gamma=0$ indicates instantaneous equilibrium with solar radiation throughout the surface, while $\Gamma=\infty$ implies a surface with a constant temperature \citepads{1989Icar...78..337S}. 

The thermal inertia plays an important role in geophysics and planetary science. It is the prime intermediary between remote temperature observations and the geological information including the particle size, rock abundance, bedrock outcropping and the degree of induration \citepads{1981Icar...45..415P,2010GeoRL..37.5404W} which in turn determine the Yarkovsky effect. 

In general, the thermal inertia is supposed to dictate the strength of the Yarkovsky effect \citepads{2015aste.book..107D}. However, results deduced from the linear model show that this is true only for the asteroids of large sizes \citepads{2020MNRAS.493.1447X}. Here we verify the statement with numerical simulations. 

\begin{figure}[htbp]
	\centering
	\resizebox{\hsize}{!}{\includegraphics{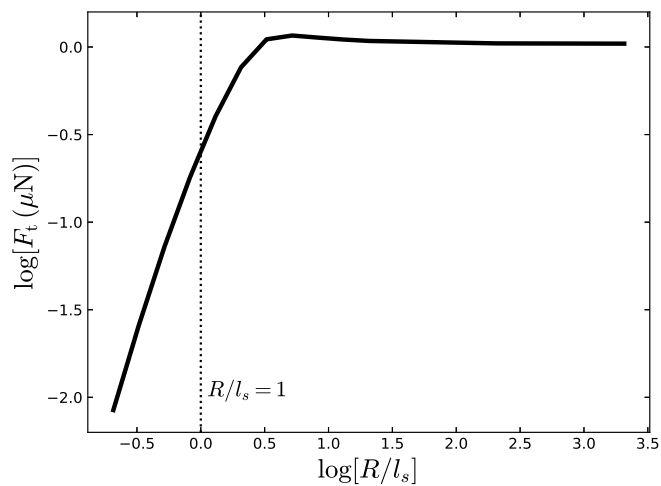}}
	\caption{The transverse Yarkovsky force of asteroids with given $\Gamma$ but different sets of $\rho, C, K$ (see text). The dotted line refers to $R/l_s=1$.
	}
	\label{fig:inertia}
\end{figure}

For a typical regolith-covered asteroid, $\rho=1500\,\text{kg\,m}^{-3}$, $C= 680\, \text{J\,kg}^{-1}\text{K}^{-1}$, $K= 0.0015\,\text{W\,m}^{-1}\text{K}^{-1}$, thus the thermal inertia $\Gamma_0=39.11\,\text{J\,m}^{-2}\text{K}^{-1}\text{s}^{-1/2}$. Keep this  $\Gamma_0$ constant, we vary $K$ and $\rho C$ according to $K=\Gamma_0^2/\rho C$, and make numerical simulations of the Yarkovsky effect. In addition, from the definition of penetration depth, we have $l_s=\sqrt{P/2\pi} K \Gamma$, thus a varying $K$ (but fixed $\Gamma$) can be translated to a varying $R/l_s$, where $R$ is the fixed radius of the asteroid ($1.0$\,m here). We show the transverse Yarkovsky force of an asteroid with constant thermal inertia ($\Gamma_0$) but varying $K$ in Fig.~\ref{fig:inertia}. 

For asteroids of large size ($R \gg l_s$), the Yarkovsky force remains constant, implying that it is determined (solely) by the thermal inertia $\Gamma$. For small asteroids whose sizes are comparable to or smaller than the penetration depth, however, the force varies even the same $\Gamma$ value is kept,  implying that the $\Gamma$ solely is not sufficient to determine the strength of Yarkovsky effect. In fact, the solution in which the heat transfer depends only on $\Gamma$  can be obtained only when an adiabatic boundary condition was fulfilled \citepads{1989Icar...78..337S}, i.e. a deep layer of constant temperature exists.  When the asteroid is so small, or equivalently the thermal conductivity $K$ is so large, that its radius is comparable to the penetration depth, which in practice is rare, the Yarkovsky effect cannot be correctly estimated if only the thermal inertia is known.  

\subsubsection{Dependence on $r$}
\begin{figure}[htbp]
	\centering
	\resizebox{\hsize}{!}{\includegraphics{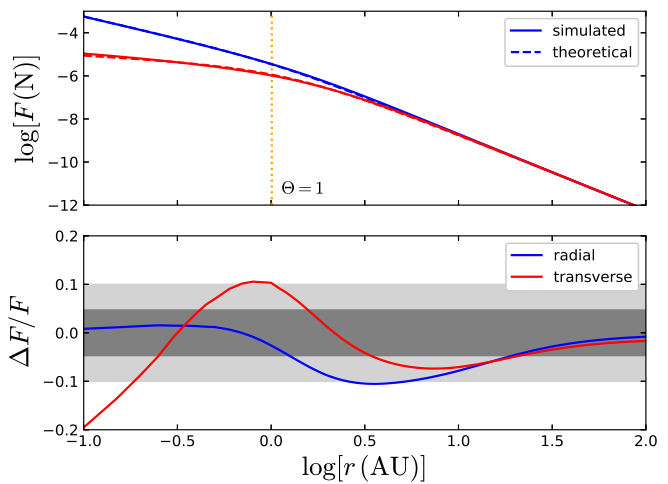}}
	\caption{Same as Fig.~\ref{fig:K} but for varying heliocentric distance $r$.
	}
	\label{fig:r}
\end{figure}

As the original drive of the Yarkovsky force, the solar radiation flux is inversely proportional to the square of heliocentric distance $r$, thus the Yarkovsky effect changes with respect to $r$. Fix other parameters at typical values but vary the heliocentric distance, we make the similar calculations and comparisons as in previous subsections and summarize the results in Fig.~\ref{fig:r}. 

The Yarkovsky effect declines quickly as the heliocentric distance $r$ increases. Further away from the Sun, the equilibrium temperature on an asteroid decreases, thus the thermal parameter $\Theta$ in Eq.~\eqref{eq:theta} increases. The upper panel of Fig.~\ref{fig:r} can be divided into two regimes. In the left regime of small $r$ and $\Theta<1$, the transverse force is smaller than the radial force. To the right of $\Theta=1$, the Yarkovsky effect is relatively weak but the radial and transverse forces are almost equal to each other. 
 
The lower panel of Fig.~\ref{fig:r} shows that the relative difference of the forces obtained from numerical simulations and derived from the linear model is within 10\% in most cases and converges to null when $r$ increases. However, the consistency is broken when the asteroid is close to the Sun, with the difference between the transverse forces derived from two methods reaching 20\% at $r = 0.1$\,AU. Therefore, particular attention should be paid if we apply the linear model to asteroids on orbit close to the Sun, or equivalently if the central star has a high luminosity.

\subsection{Biaxial ellipsoid}

\citetads{1998A&A...338..353V} also developed a linear model for biaxial ellipsoids rotating around the axis of symmetry. They considered a class of biaxial ellipsoids with different equatorial radii $R$ and polar radii $e_1 R$, and compared their Yarkovsky forces with those of spherical bodies with the same volume.  

\begin{figure}[htbp]
	\centering
	\resizebox{\hsize}{!}{\includegraphics{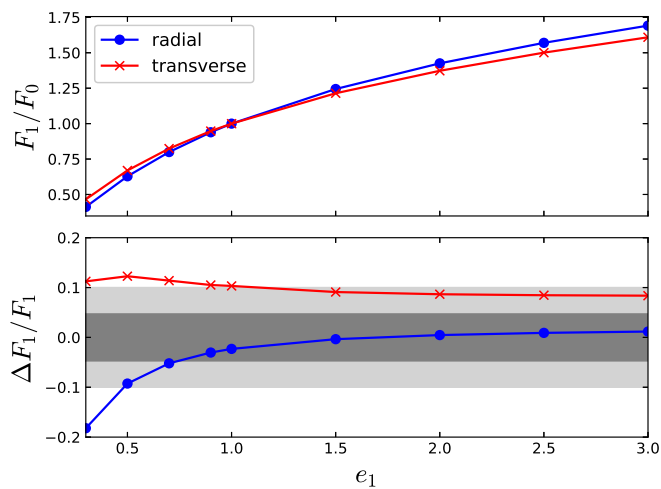}}
	\caption{The Yarkovsky forces acting on biaxial ellipsoids and the difference between the linear model and numerical simulations. Upper panel: The ratio of the forces acting on ellipsoids $F_1$ to those on the sphere $F_0$ versus oblateness $e_1$. Lower panel: The relative differences between the forces derived from linear model and numerical simulations. Blue and red are for the radial and the transverse components, respectively.
	}
	\label{fig:2d}
\end{figure}

Keeping the physical and dynamical parameters the same as the ones adopted for the spherical asteroid model in previous subsection, we make several calculations of the Yarkovsky effect for ellipsoids with the same volume as that of spherical asteroid of  radius $1.0$\,m but with oblateness $e_1$ ranging from 0.3 to 3.0. The results are summarized in Fig.~\ref{fig:2d}. We note that the rotation around the polar axis for ellipsoids with $e_1>1$ is unstable and thus rare in reality. Our main concern  in this paper is the calculation of the Yarkovsky effect.

In Fig.~\ref{fig:2d}, the Yarkovsky force acting on biaxial ellipsoids is denoted by $F_1$ and that on the same-volume sphere by $F_0$. The ratio $F_1/F_0$, both for the radial and transverse components, increases from $\sim$0.5 to $\sim$1.75 when $e_1$ increases. For oblate ellipsoids ($e_1<1$), the Yarkovsky forces are smaller because the normal vectors of facets on the surface are more likely to be aligned with the spin axis. On the contrary, the forces are relatively larger for prolate ellipsoids ($e_1>1$). 

The lower panel of Fig.~\ref{fig:2d} shows the relative differences between the results for biaxial ellipsoids derived by the linear model and numerical simulations. Although in most cases the deviation ($\Delta F_1$) of the linear model from the numerical simulation ($F_1$) is within 10\% (for transverse force) or even 5\% (for radial force), it should be noted that the transverse force derived from the linear model is always larger than the numerical results ($\Delta F_1 >0$), implying a slight but systematic overestimation exists in the linear model for the semimajor axis drift driven by the Yarkovsky effect. We also note that for oblate ellipsoids the difference goes up with decreasing oblateness parameter $e_1$, and it is worse for the radial force (up to $\sim$20\% when $e_1\rightarrow 0.3$). In fact, the temperature variation around the equatorial area is relatively large for oblate ellipsoids, so that the linear approximation causes bigger deviation for the oblate ellipsoids than for prolate ellipsoids.

\subsection{Triaxial ellipsoid}
Releasing the assumption of having an axis of symmetry from the biaxial ellipsoid model, we have a triaxial ellipsoid that could serve as a better approximation to many asteroids, and there is no any longer the linear model of Yarkovsky effect available for this case. The ellipsoids considered here have the same volume as a sphere of radius $1.0$\,m, and they are characterized by three free sem-axes ($a_\mathrm{X}$, $a_\mathrm{Y}$, $a_\mathrm{Z}$). Two parameters $e_1, e_2$ describing the shape are defined as: 
	\begin{equation}
	e_1=\frac{a_\mathrm{Z}}{\sqrt{a_\mathrm{X}a_\mathrm{Y}}}, \hspace{0.5cm} e_2=\frac{a_\mathrm{X}}{a_\mathrm{Y}}\,.
	\end{equation}
Obviously, when $e_2=1$ a triaxial ellipsoid reduces to a biaxial ellipsoid. 
And we note that ellipsoids characterized by parameter $e_2$ and $1/e_2$ are the same in shape. 

\begin{figure}[htbp]
	\centering
	\resizebox{\hsize}{!}{\includegraphics{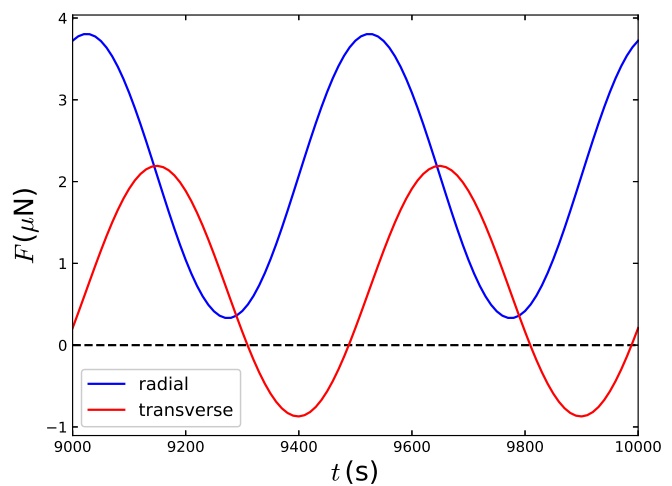}}
	\caption{Evolutions of the radial force (blue) and transverse force (red) acting on a triaxial ellipsoid in one rotation period. The flattening paramters $e_1=0.3$ and $e_2=3$. 
	}
	\label{fig:3d_example}
\end{figure}

Without the rotational symmetry, the temperature distribution on a triaxial ellipsoid's surface in the non-rotating frame should be periodic rather than static, thus the Yarkovsky force most probably is not constant, as the example in Fig.~\ref{fig:3d_example} shows. Both the radial and transverse forces oscillate with a period of half of the rotation period. And we find that the oscillation amplitudes will be amplified if $e_2$ increases. During the oscillation, the transverse force may even change direction (from positive to negative, or \textit{vice versa}). 

As the net recoil force of the thermal radiation, the Yarkovsky force basically points to  the opposite of the surface normal direction at the hottest spot. Thus, the angle $\theta$ between the radial-in vector (pointing from subsolar point to the Sun) and the surface normal vector at the hottest spot determines the direction of the Yarkovsky force. Apparently, for a prograde (retrograde) spherical asteroid, $\theta >0$ ($\theta <0$) and the transverse Yarkovsky force is positive (negative). But for a prograde rotating (around $Z$-axis) triaxial ellipsoid, when the long axis ($X$-axis for $e_2>1$) just overpasses the subsolar point, we may have $\theta < 0$ for a while, of which the duration depends on $e_2$ and thermal parameters. This is a simple explanation of why the curve of transverse Yarkovsky force periodically crosses $F=0$ in Fig.~\ref{fig:3d_example}. 

\begin{figure}[htbp]
   \centering
   \resizebox{\hsize}{!}{\includegraphics{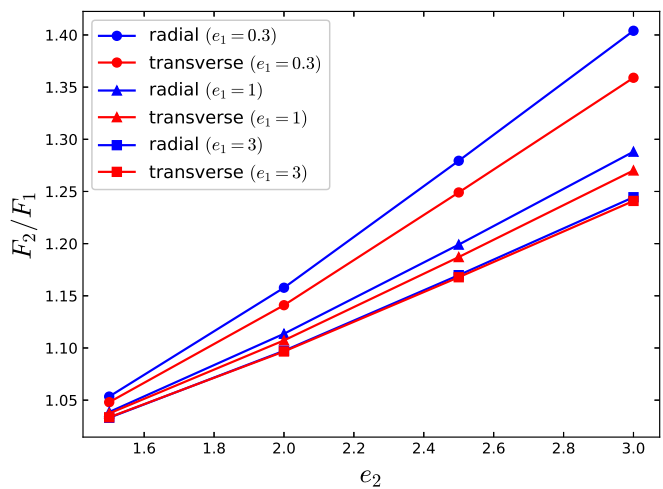}}
      \caption{Ratio of the forces acting on triaxial ellipsoids $F_2$ to those on biaxial ellipsoids of the same volume $F_1$ versus the flattening parameters $e_1$ and $e_2$. Blue and red are for radial and transverse forces, while solid circles, triangles and squares refer to $e_1=0.3, 1$ and $3$, respectively. 
                       }
         \label{fig:3d}
   \end{figure}

The oscillating force can be averaged over one rotation period. To compare the Yarkovsky effect on the triaxial ellipsoid to biaxial ellipsoid, we calculate the time-averaged forces (denoted by $F_2$) for several triaxial ellipsoids, of which the volume, $e_1$ value, and thermal parameters are the same as the corresponding biaxial ellipsoids. We denote the forces on the biaxial ellipsoids by $F_1$ and plot the ratio $F_2/F_1$ in Fig.~\ref{fig:3d}. 

The results show that larger Yarkovsky forces are produced on triaxial ellipsoids than biaxial ellipsoids, and the smaller $e_1$, the larger the difference. For a given $e_1$, the ratio $F_2/F_1$ increases as $e_2$ increases. Especially for prolate ellipsoids ($e_1>1$), the Yarkovsky force on the triaxial ellipsoid is larger than the one on the biaxial ellipsoid ($F_2>F_1$), while the latter is larger  than the force on sphere of the same volume ($F_1>F_0$, see Fig.~\ref{fig:2d}). Consequently, the Yarkovsky effect on prolate ellipsoids is significantly stronger compared to the spherical asteroids. If we adopt the linear model for spheres or for biaxial ellipsoids to calculate the Yarkovsky effect on a triaxial-ellipsoid-like asteroid, the estimation may have an error of up to $\sim$40\% (Fig.~\ref{fig:3d}), depending on how much the shape deviates from a sphere or a biaxial ellipsoid. 

Another interesting inference from above calculations is about the evolution of binary asteroids. Consider a pair of asteroids that are similar in size, rotation and composition. If their shapes are different from each other, they might suffer significantly different Yarkovsky effects, resulting in the disassembly of the binary. Therefore, we are likely to see that the pair of asteroids in a binary are similar to each other in shape.

\subsection{Real asteroids}

Although a growing number of observations indicate that asteroids are generally of irregular shapes even if small-scale structures are ignored, it is common to estimate the diurnal Yarkovsky effect by assuming the asteroid as an equal-volume sphere and thus using the linear model. But on the one hand, small asteroids that usually experience the Yarkovsky effect the most more likely have irregular shapes, because their self-gravity is too weak to modify them towards symmetrical shapes unless other effects such as rotational deformation are in act. On the other hand, the linear model introduces some errors even for regular shapes like spheres or biaxial ellipsoids, as we have shown above. Instead of creating more fictitious shapes, here we use a sample of real asteroid shapes to check the difference between the approximation of linear model and numerical simulations.  

\begin{figure}[htbp]
	\centering
	\resizebox{\hsize}{!}{\includegraphics{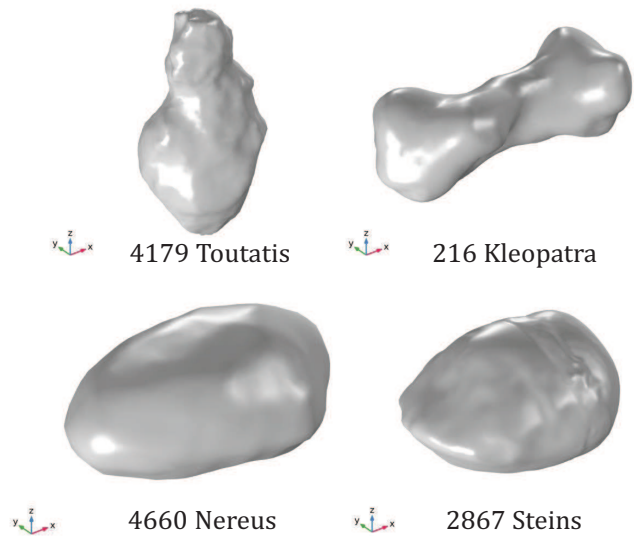}}
	\caption{The shapes of four asteroids. The spin axis goes through the barycentre (geometric centre) and is in the direction of $Z$-axis.  
	}
	\label{fig:shapexam}
\end{figure}

The shapes of 34 real asteroids adopted here are downloaded from the Planetary Data System\footnote{\url{https://sbn.psi.edu/pds/shape-models/}}, and we show 4 of them in Fig.~\ref{fig:shapexam}. To facilitate comparison, all asteroids are scaled to have the same volume as a 10\,m-radius sphere, and assumed to have the same thermal parameters $\rho, C$ and $K$ as . All but one (4179 Toutatis, see Fig.~\ref{fig:shapexam}) of them are assumed to rotate around the principal axis of the maximum moment of inertia with a period of 30\,min (1800\,s), and they are put on circular orbits at 1\,AU from the Sun. We emphasize here that these 34 asteroids provide us nothing else but a sample of diverse shapes. We note that the arbitrarily selected rotation period (30\,min) here is short and not common, but the purpose of this paper is not to give the Yarkovsky effect for a specific asteroid, and in addition, as we mentioned previously in Section \ref{sect:mesh}, the period can be traded with the density $\rho$ and heat capacity $C$.  Moreover, the dependence of the Yarkovsky effect on rotational period is already known \citepads[see e.g. ][]{2020MNRAS.493.1447X}.

For the sake of making direct comparisons, we calculate the rate of semimajor axis drifting ($\dif a /\dif t$) instead of the surface temperature or the Yarkovsky forces, simply because the semimajor axis drifting is the most prominent manifestation of the diurnal Yarkovsky effect. We calculate the $\dif a /\dif t$ by a semi-analytical method as follows.

In the study of diurnal Yarkovksy effect, since the rotation period is much shorter than the orbital period (1\,yr), it is reasonable to assume that the temperature distribution on the asteroid's surface has reached a (dynamical) equilibrium state at any phase of its orbit.
Therefore, it is possible to calculate the surface temperature and average the Yarkovksy force over one rotational period on each position along the orbit. Substitute the Yarkovsky force, now described by a function of the orbital phase (true anomaly), to the Gaussian planetary equations and average over one revolution, we obtain the rate of semimajor axis $\dif a/\dif t$. 

   \begin{figure}[htbp]
	\centering
	\resizebox{\hsize}{!}{\includegraphics{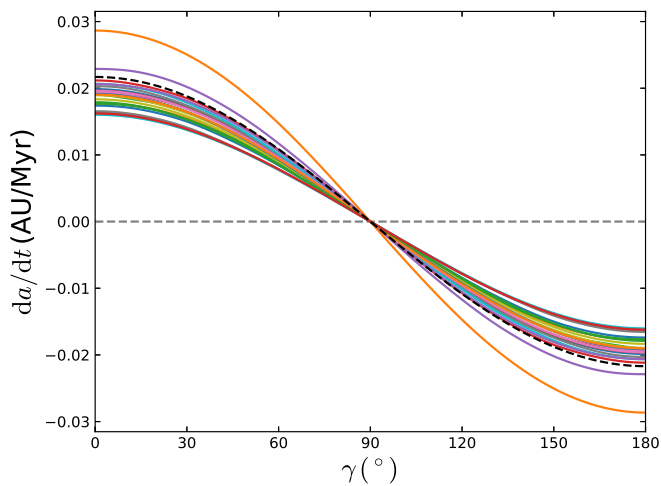}}
	\caption{The rate of semimajor axis drifting $\dif a/\dif t$ versus the obliquity of spin axis $\gamma$. Solid curves represent simulated results for 34 asteroids of 10\,m radius with real shapes while the dashed curve represents the analytical result from the linear model for a 10\,m-radius sphere. 
	}
	\label{fig:shapes}
\end{figure}

We summarize our calculations in Fig.~\ref{fig:shapes}, in which the obliquity of spin axis $\gamma$ of each asteroid is tested in the range of from $0^\circ$ to $180^\circ$.
In the linear model for a spherical asteroid, $\dif a/\dif t \propto \cos\gamma$, just as the dashed curve in Fig.~\ref{fig:shapes} shows, and it seems that this relation holds for asteroids of arbitrary shapes. For all asteroids in the sample, the diurnal Yarkovsky effect drives the prograde-spinning asteroids ($\gamma<90^\circ$) to migrate outward ($\dif a/\dif t>0$) and the retrograde rotators ($\gamma>90^\circ$) inward ($\dif a/\dif t<0$). The dependence of migration direction on obliquity is not affected by the shape. 

However, the drifting rate $|\dif a/\dif t|$ is moderately affected. Except for two ``outliers'' corresponding to Toutatis-like and Kleopatra-like asteroids (see Fig.~\ref{fig:shapexam}), all the rest 32 asteroids of real shapes migrate in slower paces than that analytically derived from the linear model for spherical asteroids (dashed curve in Fig.~\ref{fig:shapes}). At $\gamma=0^\circ$, the linear model gives $\dif a/\dif t= 0.0217$\,AU/Myr, while the numerical results of the 32 asteroids distribute from 0.0160\,AU/Myr to 0.0212\,AU/Myr, with the average value of 0.0187\,AU/Myr. Such statistical result reveals the possible error of the linear model. If we adopt the linear model for spherical asteroids to calculate the semimajor axis drift of asteroids with these real shapes, we will overestimate $\dif a/\dif t$ 
averagely by 16\%, while the maximum deviation can reach 35\%. 

The slowest Yarkovsky migration (0.0160\,AU/Myr) is found to happen on the Nereus-like asteroid, and the second slowest one is Steins-like, both shown in Fig.~\ref{fig:shapexam}. Compared to other asteroids, they are apparently more oblate than others. On an oblate asteroid, the normal directions of surface elements have relatively small components parallel to the equatorial plane, thus the recoil force of the thermal radiation (always in the normal direction) is relatively small and the migration is relatively slow. On the contrary, if the normal direction of surface elements tends to be parallel to the equatorial plane, a relatively large migration speed will be expected. We note that the two ``outliers'' of faster migration in Fig.~\ref{fig:shapes} are the Toutatis-like asteroid ($\dif a/\dif t=0.0287$\,AU/Myr) and the Kleopatra-like asteroid ($\dif a/\dif t=0.0229$\,AU/Myr). They do have more surface elements facing outwards in direction parallel to the equatorial plane (see Fig.~\ref{fig:shapexam}). Particularly, in our calculations, 4179 Toutatis is simply assumed to rotate around its long axis, although it is in fact in a complex rotation state \citepads[see e.g.][]{2015MNRAS.450.3620Z}.

In fact, an index of shape can be defined as follows to tell directly the strength of the Yarkovsky effect in driving the semimajor axis drifting. 

The shape of an asteroid can be uniquely described by a set of surface elements with different sizes $\dif S$ and  orientations $\mathbf{n}$, while the combination of ($\dif S$, $\mathbf{n}$) determines the capability of thermal re-radiation in a certain direction, i.e., the strength of the Yarkovsky effect. 

Since the penetration depth is very small compared to the radius of the object with thermal parameters adopted in this paper ($\sim$0.5\,mm), the lateral heat transfer can be ignored and the heat transfer in a rotational period must be strongly localized. In a certain surface element, the thermodynamic process of absorption, storage and re-radiation of energy should occur mainly inside the same surface element. As a result, the Yarkovsky effect, arising from the recoil force of re-radiation, is determined by the energy budget and the orientation of each individual surface element. The overall Yarkovsky effect of an asteroid is the integration of these surface elements across the whole surface. And if we focus on the influence of shape on the Yarkovsky effect, we can use such integration as an index, and a comparison to the index of a standard model (e.g. a sphere) will give directly the relative strength of Yarkovsky effect. 

Consider a surface element $\dif\mathbf{S}$ of area $\dif S$ at its local noon, when the Sun is on the plane spanned by its normal vector $\mathbf{n}$ and the $\mathbf{Z}$-axis (spinning axis). We note that these vectors are defined in an asteroid-centred inertial frame. The $\dif\mathbf{S}$ absorbs the solar radiation with a flux (power) $E_{in}$:
\begin{equation}
 E_{in} \propto \dif \mathbf{S}\cdot (-\mathbf{n}_{in}) = -(\mathbf{n} \cdot \mathbf{n}_{in})\dif S,
\end{equation} 
where $\mathbf{n}_{in}$ is the direction of incident radiation, lying in the equatorial plane ($\mathbf{XY}$ plane) when $\gamma=0$. All the absorbed energy will be re-radiated out from the same area (localization assumption) after a delay which, depending on thermal inertia $\Gamma$, is constant for each surface element. Correspondingly, the recoil force of the re-radiation will be at the reverse normal direction $(-\mathbf{n})$ when $\dif\mathbf{S}$ reaches the peak temperature, and the deviation from the local meridian normal direction is determined by the thermal delay and the object's rotation period $P$. In addition, considering the fact that the thermal evolution of a surface element during a rotational period is (approximately) symmetric with respect to the peak temperature position, a simple calculation reveals that the overall recoil force produced by this surface element during a rotational period is proportional to the force when it is experiencing the peak temperature (at the same direction for sure). 

Since all surface elements have the same $\Gamma$ and $P$, a constant delay between the absorption and re-radiation of energy for each surface element indicates that the projection of the overall recoil force of each surface element over a rotational period on the $\mathbf{XY}$ plane points to the same direction. An integration of all these surface elements across the whole surface of an object leads to the overall Yarkovsky force, the projection of which on $\mathbf{XY}$ plane surely points to the same direction too. Therefore, when we discuss the relative strength of Yarkovsky effect among asteroids with the same $\Gamma$ and $P$, we know that they all have the same delay in direction and the specific value of delay does not matter. Without loss of generality, the delay is assumed to be null, i.e. the re-radiation is assumed to happen simultaneously as the absorption in the direction $\mathbf{n}$ at noon. 

In this paper, we investigate the semimajor axis migration due to the Yarkovsky effect, thus only the force component in the orbital plane (coinciding with the equatorial $\mathbf{XY}$ plane when $\gamma=0^\circ$) is of interest. Finally, for null delay, the re-radiation power from $\mathrm{d}\mathbf{S}$ that produces recoil force on the orbital plane is 
\begin{equation}
E_{out} \propto \left[\mathbf{n} \cdot (-\mathbf{n}_{in})\right] E_{in} \propto (\mathbf{n} \cdot \mathbf{n}_{in})^2 \dif S .
\end{equation}
The total effect can be obtained by integrating $(\mathbf{n} \cdot \mathbf{n}_{in})^2 \mathrm{d}S$ over the whole surface of an object. We note that this integration should be done by rotating the object in the asteroid-centered inertial frame, and the result is an ``effective area''. As a ``standard shape'', the effective area of a sphere with radius $R$ is $8\pi R^2/3$. 

\begin{figure}[htbp]
	\centering
	\resizebox{\hsize}{!}{\includegraphics{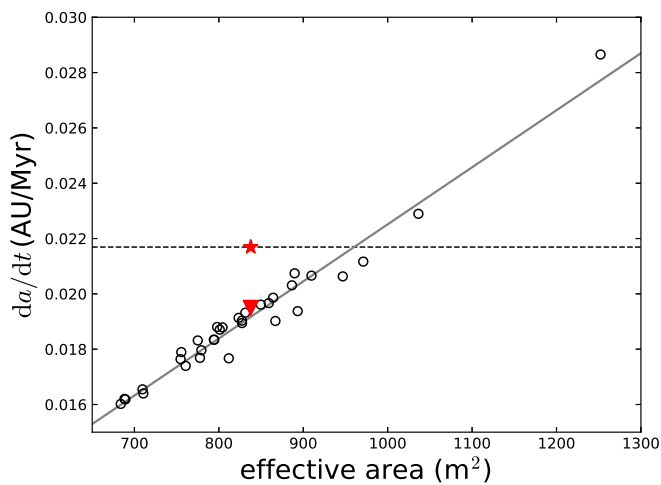}}
	\caption{The semimajor axis drifting rate versus the effective area (see text). The red solid triangle stands for the numerical results of a spherical asteroid and the red star is from the linear model for the same asteroid. The solid line is the least squares fit of the results of 34 shapes.
	}
	\label{fig:shapeind}
\end{figure}

The effective area tells the capability of an object to produce the recoil force by re-radiation, thus can serve as the index of relative strength of the Yarkovsky effects acting on asteroids of different shapes but the same thermal parameters. Again, use the 34 real asteroids (all scaled to have the volume of a 10\,m-radius sphere) as examples, we calculate their effective areas, and we show in Fig.~\ref{fig:shapeind} the relation between the $\dif a/\dif t$ of these asteroids with obliquity $\gamma=0^\circ$ and their effective areas. The $\dif a/\dif t$ values obtained from the numerical simulation and derived from the linear model for the asteroid of standard shape (10\,m-radius sphere, with an effective area of 838\,m$^2$) are also plotted for reference. 

Just as expected, an apparent linear relationship for all numerical results can be seen in Fig.~\ref{fig:shapeind}, even the two ``outliers'' in Fig.~\ref{fig:shapes} (Kleopatra-like and Toutatis-like asteroids) are found to tightly sit on the line. Since the linear model overestimates the Yarkovsky effect, the result derived from it (the red star) deviates from the linear relation. Corresponding to the widely varying shapes, the effective areas of the 34 asteroids in the sample spread from 684\,m$^2$ (Stein-like) to 1252\,m$^2$ (Toutatis-like), and all of them follow the same linear relation very well, indicating that this relationship is robust. 

All asteroids have been assumed to rotate around $Z$-axis which makes the Tuotatis-like asteroid very efficient in producing the Yarkovsky force since it has the longest axis in $Z$-direction (and thus the largest effect area). As shown in Fig.~\ref{fig:shapexam}, the shape of Toutatis is more or less like a triaxial ellipsoid with axes satisfying $a_Z\gg a_X\approx a_Y$. If such an asteroid rotates around its short axes, the sharp ends of the long axis would cause the transverse force (thus $\dif a/\dif t$) to change from positive to negative when the long axis overpasses the subsolar point (see Fig.~\ref{fig:3d_example} and explanation there). However, in calculating the effective area, each surface element has been simply assumed to provide a contribution always in a same direction. Therefore, if a Tuotatis-like asteroid rotates around the short axes, the Yarkovsky effect will be overestimated, and the corresponding point would deviate from (and locate below) the fitting line in Fig.~\ref{fig:shapeind}. 

In fact, Toutatis is a tumbling asteroid. It rotates basically around its long axis which is simultaneously precessing with a comparable period \citepads{2015MNRAS.450.3620Z}. Thus its angular velocity vector $\vec{\omega}$ describes a cone whose axis tilts by angle $\alpha$ from the normal of the orbital plane and whose aperture is $2\beta$ (thus the precession amplitude is $\beta$). Our preliminary calculations show that the average Yarkovsky effect is proportional to $\cos\alpha\cos\beta$, therefore the corresponding effective area of such a tumbler can be defined as the original effective area (as principal rotator around $Z$-axis) multiplied by $\cos\alpha\cos\beta$. We note that as $\beta\rightarrow 0$, this tumbler reduces to a principal rotator with spin obliquity $\gamma=\alpha$, and the Yarkovsky drift $\dif a/\dif t \propto \cos\gamma$, as both the linear theory and numerical simulations agree. However, further investigation on the effective area for asteroids in complicated rotation states is needed. 

We emphasize that the effective area is independent of the thermal physical process, but   only determined by the asteroid’s shape. With this relationship, we are now be able to obtain directly the Yarkovsky effect of any irregular asteroid by comparing its effective area to the one of standard shape (sphere), free from complicated and expensive calculations of temperature on the complex surface.

\section{Conclusions} \label{sect:conclusion}
In the past three decades, observations have been showing the importance of the Yarkovsky effect on the motions of small celestial bodies. 
The recoil force of thermal irradiation may modify the rotation and orbit of asteroids, and the semimajor axis drifting is the most prominent long-term effect. 

The linear model proposed by \citetads{1998A&A...335.1093V,1998A&A...338..353V} has been widely adopted and successfully produced good estimations for Yarkovsky effects on asteroid populations. In the linear model, the temperature variation on the asteroid surface ($\Delta T$) is assumed to be so small that $(\bar{T}+\Delta T)^4$ can be approximated by $\bar{T}^4+4\Delta T$, and the shape of asteroid is supposed to be spherical or close to a sphere. 
However, these two conditions could be deviated in reality. In this paper, using a multiphysics software COMSOL, we numerically calculate the diurnal Yarkovsky effect in three dimensions and make comparisons with the results derived from the linear model. 

In the calculations, we find that the radiation flux from the Sun introduces temperature fluctuation only into a thin layer under the surface of the asteroid. Depending on the thermal conductivity, the thickness of this thermally-affected-layer, beneath which the temperature is constant, is about ten times the penetration depth. Therefore, the numerical calculation of surface temperature that determines the Yarkovsky effect can be performed only in this subsurface layer, that is, the asteroid can be regarded as the combination of a thermally affected shell and an isothermal core. 

After obtaining numerically the temperature distribution, we calculate the Yarkovsky  force or the semimajor axis drifting rate. 
Comparisons with results derived from the linear model prove that the linear model for spherical bodies works very well, with the relative errors being within 10\% in most cases. However, the linear model could moderately deviate from the real situation when the asteroid conducts heat poorly (small conductivity) or it is very close to the Sun. 

The thermal inertia $\Gamma$ indicates the capacity to store energy and to delay the transmission of heat in the material. We find that $\Gamma$ dictates the strength of diurnal Yarkovsky effect for large bodies. But when the asteroid is so small (or equivalently the thermal conductivity is so large) that its size is comparable to the penetration depth, the Yarkovsky effect cannot be solely decided by $\Gamma$. Another thermal parameter $\Theta$, that specifies the body's ability of retaining temperature changes over a rotation period, may control the delay angle between the subsolar point and the hottest spot on the surface thus determines the ratio between radial and transverse components of the Yarkovsky force. 

Compared to sphere, a prolate biaxial ellipsoid produces stronger diurnal effect while an oblate ellipsoid has a weaker effect. The results derived from the linear model for biaxial ellipsoids are overall consistent with numerical simulations. But a systematic overestimation of transverse Yarkovsky force can be found in the linear model, and the deviation is greater for oblate ellipsoids.  

For triaxial ellipsoids, the diurnal Yarkovsky effect turns to be a periodic effect, for which no theoretical solution is available now. Within one rotation period, the transverse Yarkovsky force could change direction, from positive to negative and \textit{vice versa}. Our numerical results show that a triaxial ellipsoid always receives a stronger average diurnal Yarkovsky effect than that acting on the biaxial ellipsoid with the same volume and flattening parameter $e_1$. When a linear model is used to calculate the effect, the deviation from numerical results may reach 40\%, depending on the flattening parameters. 

Apart from these fictitious regular shapes, we have also considered 34 real shapes, many of which are extremely irregular. Our calculations show that the semimajor axis drifting rates tend to be overestimated by 16\% on average and up to 35\% if the linear model for spheres is applied. We introduce the ``effective area'' to quantify the relative diurnal Yarkovsky effect on irregularly-shaped asteroids, and we find a perfect linear relationship between the semimajor axis drifting rate and the effective area. For an asteroid of any shape, by comparing its effective area with that of a standard model (e.g. a spherical asteroid), the Yarkovsky effect can be obtained directly without calculating the temperature distribution on the surface assuming the same thermal parameters. Combined with our analysis of the deviation of linear model for Yarkovsky effect, the effective area provides an efficient approach to accurately calculate the Yarkovsky effect for asteroids of irregular shapes analytically.

The effective area defined here reflects only the geometrical property of the surface, and the shadowing effect has been neglected, and we also assume that the heat transfer is localized. These assumptions are reasonable for the majority of real asteroids, given the penetration depth is much smaller than the radius. However, the effective area for asteroids of extremely irregular shape and/or of complicated rotation needs to be improved. 
Similarly, an ``effective torque'' can also be defined to estimate the YORP effect, which will be discussed in a separate paper.


\begin{acknowledgements}
	We thank the anonymous referee for the valuable comments. This work has been supported by the National Key R\&D Program of China (2019YFA0706601) and National Natural Science Foundation of China (NSFC, Grants No.11933001 \& No.11473016). We also acknowledge the science research grants from the China Manned Space Project with NO.CMS-CSST-2021-B08. JYL’s contribution to this work is not supported by any funds. The temperature distribution on the asteroid's surface is computed using  COMSOL software. 
\end{acknowledgements}

%
%

\end{document}